\documentclass[preprint]{aastex}
\def\ifundefined#1{\expandafter\ifx\csname#1\endcsname\relax}

\newif\ifpdf
\ifx\pdfoutput\undefined
   \pdffalse      
\else
   \pdfoutput=1   
   \pdftrue
\fi

\def\la{\mathrel{\hbox{\rlap{\hbox{\lower4pt\hbox{$\sim$}}}\hbox{$<$}}}}
\def\ga{\mathrel{\hbox{\rlap{\hbox{\lower4pt\hbox{$\sim$}}}\hbox{$>$}}}}

\newcommand{\be}{\begin{eqnarray}}
\newcommand{\ee}{\end{eqnarray}}

\ifundefined{ensuremath}\def\ensuremath#1{\relax\ifmmode{#1}}
\else${#1}$\fi\else\relax\fi
\ifundefined{nuc}\def\nuc#1#2{\relax\ifmmode{}^{#1}{\protect\text{#2}}
\else${}^{#1}$#2\fi}\else\relax\fi

\newcommand{\gcm}{g~cm$^{-3}$}
\newcommand{\kmps}{km~s$^{-1}$}

\newcommand{\msol}{\ensuremath{{\textrm{M}_\odot}}}

\def\ang{\hbox{\AA}}
\def\Tmod{\ensuremath{T_{\textrm{model}}}}

\def\tstd{\ensuremath{\tau_{\textrm{std}}}}

\newcommand{\phx}{\texttt{PHOENIX}}

\begin{document}

\bibliographystyle{apj}

\title{Analysis of the Type IIn Supernova 1998S: Effects of
Circumstellar Interaction on Observed Spectra}

\author{{Eric~J.~Lentz\altaffilmark{1}}, {E.~Baron\altaffilmark{1}}, {Peter Lundqvist\altaffilmark{2}},
{David~Branch\altaffilmark{1}}, {Peter~H. Hauschildt\altaffilmark{3}},
 {Claes Fransson\altaffilmark{2}},
 {Peter Garnavich\altaffilmark{4}}, 
{Nate Bastian\altaffilmark{1,5}},
{Alexei~V.~Filippenko\altaffilmark{6}}, {R.~P.~Kirshner\altaffilmark{7}},
{P.~M.~Challis\altaffilmark{7}}, {S.~Jha\altaffilmark{7}},
{Bruno Leibundgut\altaffilmark{8}}, {R. McCray\altaffilmark{9}}, 
{E. Michael \altaffilmark{9}},
{Nino Panagia\altaffilmark{10}}, {M.~M.~Phillips\altaffilmark{11}},
{C.~S.~J.~Pun\altaffilmark{12}}, {Brian Schmidt\altaffilmark{13}},
{George Sonneborn\altaffilmark{12}}, 
{N.~B.~Suntzeff\altaffilmark{14}}, 
{L.~Wang\altaffilmark{15}}, and {J.~C.~Wheeler\altaffilmark{15}}}

\altaffiltext{1}{Department of Physics and Astronomy, University of
Oklahoma, 440 W. Brooks, Norman, OK~73019-0261}

\altaffiltext{2}{Stockholm Observatory, SE--133~36 Saltsj\"obaden,
Sweden}

\altaffiltext{3}{Department of Physics and Astronomy \& Center for
Simulational Physics, 
University of Georgia, Athens, GA 30602}

\altaffiltext{4}{Dept. of Physics, Univ. of Notre Dame, 225 Nieuwland
Science Hall, Notre Dame, IN 45656}

\altaffiltext{5}{Dept. of Astronomy, Univ. of Wisconsin, 
475 N. Charter St., Madison, WI 53706}

\altaffiltext{6}{Department of Astronomy, University of California,
Berkeley, CA~94720--3411}

\altaffiltext{7}{Harvard--Smithsonian Center for Astrophysics,
60~Garden St., Cambridge, MA~02138}


\altaffiltext{8}{European Southern Observatory,
Karl-Schwarzschild-Strasse 2, D-85748 Garching, Germany}

\altaffiltext{9}{JILA,  Univ. of Colorado, 
Boulder, CO 80309}

\altaffiltext{10}{Space Telescope Science Institute, 3700 San Martin
Drive, Baltimore, MD~21218 (on assignment from Space Science
Department of ESA)}

\altaffiltext{11}{Carnegie Inst. of Washington,
                        Las Campanas Obs.,
                        Casilla 601, Chile}

\altaffiltext{12}{Laboratory for Astronomy and Solar Physics,
NASA/GSFC, Code~681, Greenbelt, MD~20771}

\altaffiltext{13}{Mount Stromlo Obs, Australian National Univ.
			Private Bag,
			Weston Creek P.O, ACT 2611, Australia}

\altaffiltext{14}{CTIO, NOAO, Casilla~603, La~Serena, Chile}

\altaffiltext{15}{Department of Astronomy, University of Texas,
Austin, TX~78712}

\begin{abstract}
We present spectral analysis of early observations of the Type~IIn
supernova 1998S using the general non-local thermodynamic equilibrium
atmosphere code {\tt PHOENIX}.  We model both the underlying
supernova spectrum and the overlying circumstellar interaction region
and produce spectra in good agreement with observations. 
The early spectra are well fit by lines produced primarily in the
circumstellar region itself, and later spectra are due primarily
to the supernova ejecta. Intermediate spectra are affected by both
regions.  A mass-loss rate of order $\dot M \sim
0.0001-0.001$\msol~yr$^{-1}$ is inferred for a wind speed of
$100-1000$~\kmps. We 
discuss how future self-consistent models will better clarify the
underlying progenitor structure.
\end{abstract}

\section{Introduction}
     SN 1998S was discovered on Mar. 3 UT by Zhou Wan \citep{IAUC6829}
as part of the Beijing Astronomical Observatory (BAO) Supernova Survey
\citep{IAUC6612}.  The discovery was confirmed by the Katzman
Automatic Imaging Telescope (KAIT) during the Lick Observatory
Supernova Search \citep{IAUC6627,flipper00}. SN 1998S is located in
NGC 3877, a 
spiral galaxy classified as SA, with a heliocentric velocity of
902~\kmps\ \citep{UGC73} and a Galactic extinction of $A_B=0.01$~mag
\citep{burheilred}.

Filippenko \& Moran \citep{IAUC6829} obtained a
high-resolution spectrum of SN 1998S on Mar. 4 with the Keck-1
telescope and classified SN 1998S as a Type~II supernova (SN~II) on
the basis of broad H$\alpha$ emission superposed on a featureless
continuum.  Further spectra were obtained at the Fred L.~Whipple
Observatory (FLWO) \citep{Garn98S00} and a campaign to monitor
SN~1998S in the UV from the \emph{Hubble Space Telescope} (\emph{HST})
was mounted by the Supernova INtensive Study (SINS) team. Three epochs
have been observed with \emph{HST} --- March 16, March 30, and May
13. SN~1998S is a Type~IIn supernova \citep[SN
IIn;][]{schlegel2n90}, a classification which shows  wide variations in
the spectra \citep{filarev97}, but includes narrow lines on top of an
underlying broad-line supernova spectrum. This has been taken as
strong evidence that the supernova ejecta were interacting with a
slow-moving circumstellar wind \citep{leon98S00}, probably in a
fashion similar to (but possibly more extreme than) that of SN~1979C
and SN~1980K \citep{lentz98Saas99,bao98s00}.

In order to get an initial understanding and to confirm the basic
picture of SNe~IIn as strong circumstellar interacters, a set of
parameterized models of SN 1998S was examined with the fully
relativistic, NLTE, multi-purpose, expanding atmosphere
code, {\tt PHOENIX}, \citep[cf.][and references therein]{hbjcam99}.
{\tt PHOENIX} solves the spherically symmetric radiation transport
along with the NLTE rate equations and the condition of radiative
equilibrium (including deviations due to time dependence of the
deposition of non-thermal gamma rays). 

\begin{figure}
\epsscale{0.7}
\plotone{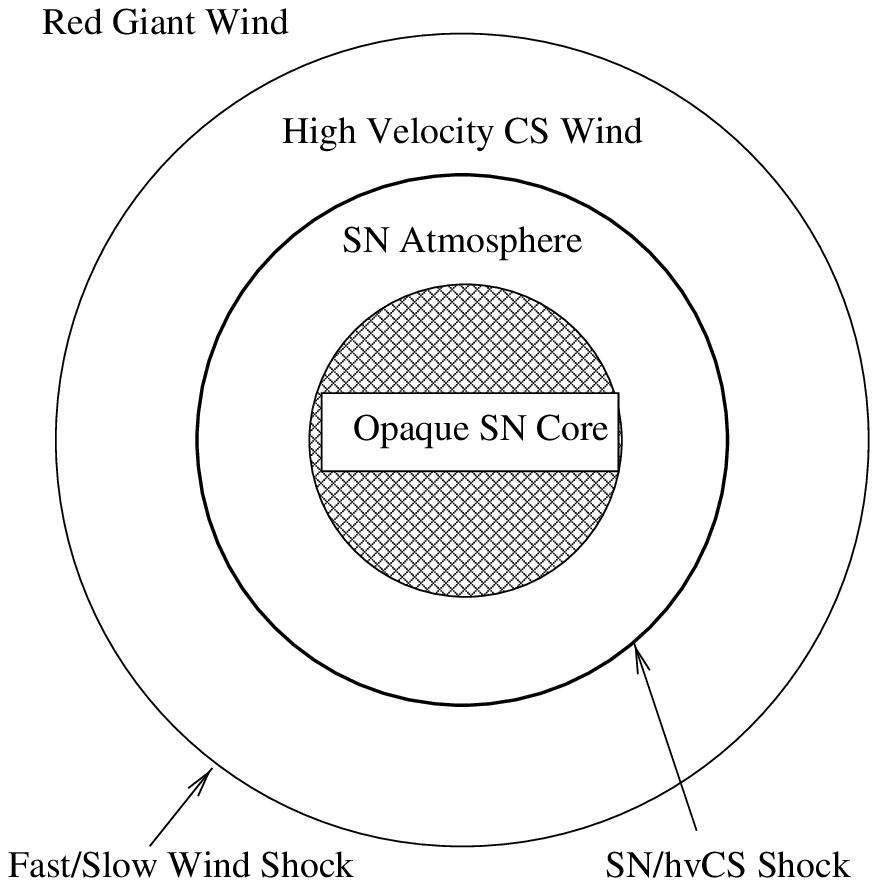}
\caption{\label{fig:diagram} Schematic diagram of SN~1998S.
hvCS stands for high velocity circumstellar material which was likely
radiatively accelerated to the high velocities seen in SN~1998S. The Red
Giant wind is assumed to have been ejected at low velocity ($\approx 10$~\kmps).}
\end{figure}

A modification of {\tt PHOENIX} is underway in order to treat
self-consistently both the underlying supernova spectrum and the
overlying circumstellar interaction region; here we treat the two regions
separately. Figure~\ref{fig:diagram} shows a schematic representation
of SN~1998S.  Some of the important coupling between the
two regions is not included in the calculations and therefore not all
of the features observed can be expected to be reproduced by the
synthetic spectra. However, our models serve to confirm the basic
picture of a SN~IIn as a Type~II supernova that interacts strongly with
a near-constant velocity wind.  We are able to identify important
physical effects that need to be included in future simulations.

While our models are spherically symmetric, \citet{leon98S00} have
shown that the spectra of SN~1998S are significantly polarized, which
could be due 
to asymmetry in the outermost SN ejecta, the circumstellar medium (CSM), or
both. \citet{gerardetal00} suggest that dust and CO are likely to have
formed in the SN ejecta while \citet{Fassia98S00} argue that
the early dust is likely to come from the CSM.

\section{Models}

In this paper we focus on three epochs in particular: an early epoch
on Mar 16,
$\sim 20$~days after explosion,
using combined \emph{HST} and ground-based spectra, where effects from
the circumstellar region dominate; Mar 30,
$\sim 34$~days after explosion,
again where there are combined \emph{HST} and ground-based data, and
where effects from both 
photospheric SN ejecta and the circumstellar region are important, and
a later ground-based spectrum from April 17, $\sim 50$ days after explosion,
where the densest circumstellar gas has been largely, but
not completely, overrun by the supernova ejecta.  In the earlier spectrum
most of the observed lines are
formed in the low-velocity circumstellar material,  whereas in the
later spectrum the lines show the
characteristic width of a Type~II supernova. A detailed analysis of
the light curve and other observed spectra will be presented elsewhere
\citep{Garn98S00}; see also \citet{leon98S00} and \citet{Fassia98S00,Fassia98S01}.

\subsection{March 16}

We have modeled the circumstellar region as a constant-velocity wind
with a density profile $\rho \propto r^{-2}$. While the underlying
radiation below the circumstellar region is in fact due to the
supernova itself and should show broad P-Cygni profiles as well as a
UV deficit due to line blanketing in the differentially expanding
supernova atmosphere, we ignore these complications for the present
discussion and assume that the underlying radiation is given by a
Planck function, with $T_{\textrm{Planck}} = 13250$~K. In future work we will
treat  the effects of the circumstellar interaction region on the
supernova itself, and couple the proper supernova boundary condition
into the circumstellar region. Nevertheless, our present 
decoupled prescription 
allows us to model the important physics, and to estimate velocities,
density profiles, and the radial extent of the circumstellar
interaction region and the supernova. The region modeled in these
calculations coincides with the region labeled ``High Velocity CS
Wind'' in Figure~\ref{fig:diagram}. High-resolution spectra
\citep{Fassia98S01} have shown that there may be several velocity
components present in the circumstellar medium with velocities as low
as 80~\kmps. We focus here on only
the higher velocity (but possibly still unresolved) components of the
CS spectrum.

\begin{figure}
\epsscale{0.8}
\plotone{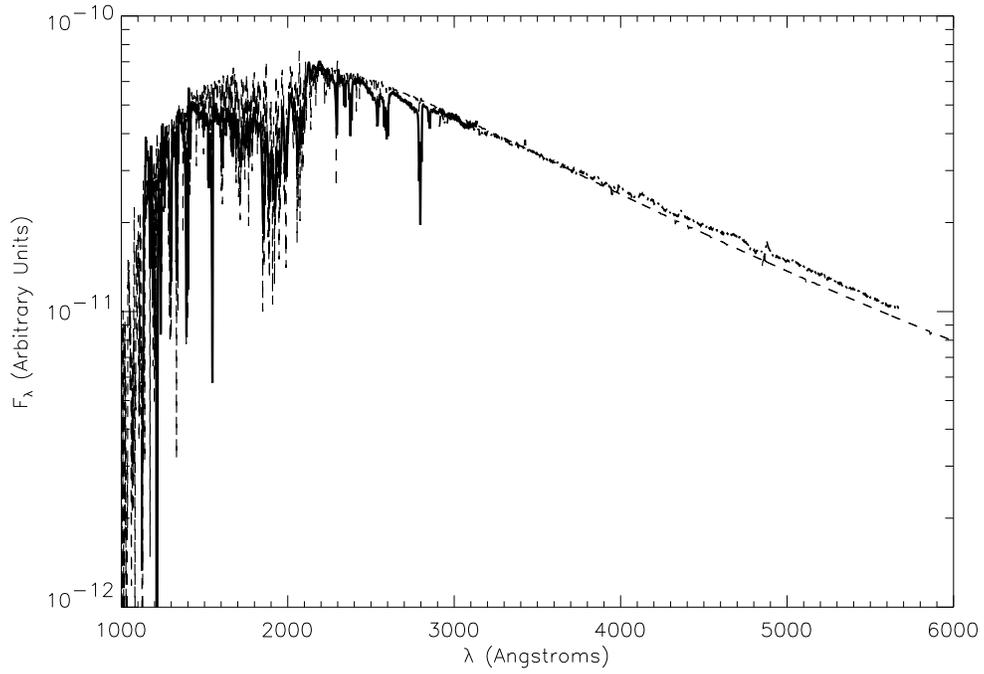}
\caption{\label{csoverview}The calculated synthetic spectrum (dashed
line) from the
circumstellar shell is compared with the \emph{HST} UV observations and the
optical spectra taken at the FLW Observatory on Mar. 16, 1998.  The
observed spectrum has been dereddened assuming $E(B-V)
=0.15$~mag and deredshifted assuming a heliocentric velocity of
902~\kmps\ in this and subsequent figures that include synthetic spectra.}
\end{figure}

Figure~\ref{csoverview} presents an overview of our best model fit
compared with the observed \emph{HST} UV + FLWO optical spectrum taken
on Mar 16, 1998. The observed spectrum has been dereddened using the
reddening law of \citet*{card89} and a color excess $E_{B-V}
=0.15$~mag \citep{Garn98S00}. 
The assumed extinction is also  in agreement with the results of
\citet{Fassia98S00} 
who find $E(B-V)=0.18 \pm 0.10$~mag. The overall agreement in the line
positions and shape of the spectrum is excellent, particularly the
pseudo-continuum near 2000~\ang. The model consists of a
constant-velocity circumstellar wind with $v_{wind} = 1000$~\kmps, an
inner density of $\rho_0 = 2.0 \times 10^{-15}$~\gcm, an inner radius
$R_{inner} = 1.0 \times 10^{15}$~cm, and an outer radius $R_{outer} =
1.5 \times 10^{15}$~cm. The total continuum optical depth at
$5000$~\ang\ (roughly the electron scattering optical depth) is $\tstd
= 0.2$, (where $\tstd$ is the total continuum optical depth at
5000~\ang), and the mass of the wind is $6\times
10^{-3}$~\msol. Assuming 
the \emph{ejected} wind velocity was 100~\kmps\ this corresponds to a
mass-loss rate of 0.0012~\msol~yr$^{-1}$,  which should be accurate to
an order of magnitude. We believe that the high velocity seen here is
due to radiative acceleration of a wind that was ejected at a lower
velocity. Since the mass-loss rate depends inversely on
the wind velocity at ejection and wind velocities typical of
red-giants are $v_{wind} \approx 10$~\kmps, we think that assuming an
ejection velocity of 100~\kmps\ allows us to estimate the mass loss
rate to an order of magnitude. 

\begin{figure}
\epsscale{0.8}
\plotone{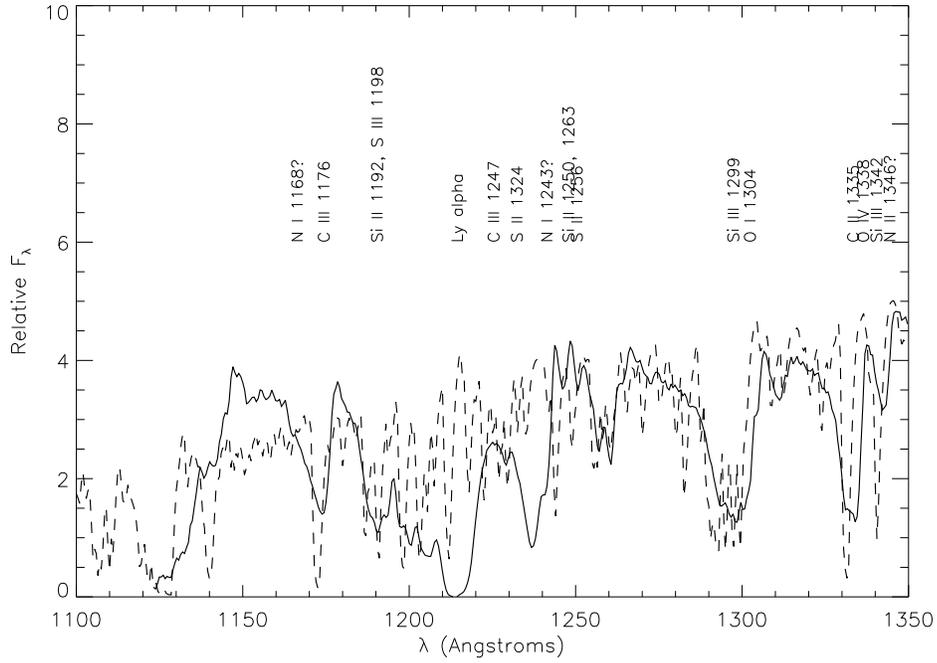}
\caption{\label{lineidfig}The calculated synthetic spectrum (dashed
line) from the
circumstellar shell is compared with the \emph{HST} UV observations of Mar.
16, 1998  and lines
are identified. Careful examination of the figure reveals that there
is an underlying broad component to the lineshapes due to the faster
supernova ejecta. However, this is not included in the present model.
}
\end{figure}

In
Figures~\ref{lineidfig}\mbox{a-f} we expand the wavelength scale and
identify the features in the observed spectrum. Several of them are
clearly pairs of interstellar absorption features where one member is
due to absorption in our Galaxy and the other is due to absorption in
the parent galaxy (Mg~II h+k shows this effect
clearly). Table~\ref{lineidtab} lists the line 
identifications.  We note that the ``interstellar'' absorption lines
in the parent galaxy may also have a circumstellar contribution. Close
examination of Figure 3 shows that the observed lines are
significantly wider than those in the synthetic 
spectrum. On the other hand, \citet{Fassia98S01} observed IR features with velocities as
low as 90~\kmps, and \citet{bowen98S00} observed UV P-Cygni features
with velocities of $\sim 100$~\kmps. Convolving the synthetic spectrum
with a Gaussian of width 400~\kmps\ improves the fit, but since we
have assumed a velocity higher than that of the lowest velocity
observed \citep{Fassia98S01}, it is difficult to separate out the
instrumental resolution ($\sim 300-400$~\kmps) from the velocity of
the circumstellar medium.  It could be that the velocity structure of
the circumstellar region is quite complicated with a higher velocity
component radiatively accelerated by the supernova, as was suggested
for SN~1993J \citep{flc96}, and a lower
velocity component  wind further away from the progenitor star.

Our model spectrum clearly does an extremely good job in reproducing
the overall shape and position of the observed features; nevertheless,
the line features are somewhat weaker in general than 
those observed. This could be due to the effects of the radiation from
the circumstellar interaction.
The effects of this radiation  are not included in these simple 
preliminary calculations. The effects of the ``top-lighting'' or
``shine-back'' are not limited to radiative transfer effects alone,
but will also affect the ionization state of the matter, particularly
if there is significant X-ray emission from a reverse shock.
In future work we will include the effects of external
irradiation from the circumstellar region and replace the simple inner
Planck function boundary condition that we have used here with a model
supernova spectrum. Such a spectrum would be hotter, but 
diluted and contain both the UV deficit of a normal Type~II
supernova as well as  broad P-Cygni features for which
there is evidence in the observed spectrum.

\subsection{March 30}

Figure~\ref{mar30data} displays the combined \emph{HST} spectra with
an optical spectrum obtained at the FLWO. It is interesting
to note that the narrow features present on Mar.~16 seem to have
disappeared, and the broad lines are all quite weak. A simple
analytical explanation of this is presented in \citet{toplight00},
which shows that with the additional emission (``toplighting'' or
``shine-back'') from the circumstellar
shell, one expects the supernova features to appear
muted. Figure~\ref{nate} displays a \phx\ spectrum,  along with the
results obtained when it has been muted
according to the prescription in \citet{toplight00}. The regular \phx\
spectrum 
is based upon the simplest assumptions: homogeneous solar
abundances, a model temperature $\Tmod = 6000$~K \citep[the model
temperature is simply a way of parameterizing the total bolometric
luminosity in the observers frame, see][]{hbjcam99}, a velocity of
5000~\kmps\ at $\tstd=1$, and a density structure $\rho \propto
r^{-8}$.  Using
Eqn.~23 of \citet{toplight00}, we have calculated the
muting, using $E=0.9$, where $E$ is the ratio of the CS intensity to that of
the SN intensity given in Eqn. 22 of \citet{toplight00}, we have
assumed a ratio of $R_{CS}/R_{Ph} =1.5$, where the ratio is the radius
of the
circumstellar shell to the  radius
of the ``SN photosphere''. While the fit is 
not terribly good, the trend is evident. Naturally, a fully consistent
model would be better, but it would require significant computational
resources to resolve both the ejecta and circumstellar
region. \citet{fransson79c84} calculated lineshapes expected from the
CS wind and the cool, dense shocked material and compared them with
those observed in SN~1979C. 

\citet{leon98S00} suggest that SN~1998S underwent a significant
mass-loss episode that ended about 60 years prior to explosion and
that there was a second, weaker mass-loss episode 7 years prior to
explosion. Thus, we may be seeing the over-running of the closest CS shell
and still observing effects of the more distant CS shells.

\subsection{April 17}

During the early evolution the nearest circumstellar material is overrun by the
supernova ejecta so the effects of the CSM on the optical and UV
spectra become smaller. Inspection of the observed optical
spectra \citep{leon98S00,Garn98S00} shows an increasing contrast in the broad
features typical of Type~II SNe during the time from the initial \emph{HST}
observation, March 16, to the FLWO spectrum of April 17.
\citet{Blaylock98S00} show that the strengthening of these
features during this transition is well reproduced by including the
effects of radiation from the circumstellar interaction region along
with the scattering of light from the supernova photosphere in the
circumstellar region.

Figure~\ref{apr17opt} displays our best model fit to the observed
optical spectrum taken at the FLWO \citep{Garn98S00}. We again use simple
assumptions: homogeneous solar
abundances, a model temperature $\Tmod = 5700$~K, a velocity of
5000~\kmps\ at $\tstd=1$, and a density structure $\rho \propto
r^{-8}$. The highest velocity in the model is only 6,000~\kmps, which
gives an indication that the ejecta are entrained by the circumstellar
material, but this is not well constrained by our models. Again,
overall the fit is very good.
The Na D line in the observed spectrum is too weak in our
synthetic spectrum, which may indicate the need to self-consistently
include the effect of the circumstellar region or may be due to
enhanced sodium. 
The extended absorption wing of H$\alpha$ is due in
our model to blending of weak Fe~II lines, although some of the
absorption may be due to Si~II. In any case it is not evidence for
high-velocity hydrogen.

\section{Conclusions}

We have shown that a simple model of an ordinary Type~II supernova
atmosphere interacting strongly with a radiatively accelerated
wind reasonably well reproduces the observed
line-widths and many of the observed features in both the UV and the
optical spectra. This model is robust in that it works well at both very early
times and more than a month after the explosion. This confirms the
general picture of SNe~IIn as being the core collapse of massive stars
that have experienced a significant mass-loss epoch and thus are
surrounded by a circumstellar medium with which the supernova ejecta
interact. As expected from our models \citep{lentz98Saas99}, SN~1998S has been
detected about 600 days after explosion at 6~cm
\citep{IAUC7322}. Although SN~1998S is about 5 times less luminous than
SN 1988Z, further monitoring of the radio light curve will be very
interesting and will help determine the mass-loss rate. From the
light curve \citet{Fassia98S00} find that the mass of the ejected
envelope was quite low and the wind was weaker than that of
SN~1988Z. SN~1998S may well be more closely related to SN~1979C and
SN~1980K.  In
future work we will develop a more self-consistent model of the
supernova circumstellar interaction, and will be able to constrain
mass-loss rates and total mass loss which are of great interest for
the theory of stellar evolution.

\acknowledgments We thank the anonymous referee for suggestions which
considerably improved the presentation of this paper. PHH was
supported in part by the P\^ole Scientifique de Mod\'elisation
Num\'erique at ENS-Lyon. NB was supported in part by an NSF REU
supplement to the Univ. of Oklahoma. This work was supported in part
by NSF grants AST-9731450, AST-9417102, AST-9987438, and AST-9417213;
by NASA grant NAG5-3505 and an IBM SUR grant to the University of
Oklahoma; by NSF grant AST-9720704, NASA ATP grant NAG 5-8425, and
LTSA grant NAG 5-3619 to the University of Georgia; and by NASA
GO--2563.001 to the SINS group from the Space Telescope Science
Institute, which is operated by AURA, Inc.~under NASA contract NAS
5--26555.  Some of the calculations presented in this paper were
performed at the San Diego Supercomputer Center (SDSC), supported by
the NSF, and at the National Energy Research Supercomputer Center
(NERSC), supported by the U.S. DOE; we thank both these institutions
for a generous allocation of computer time.  This research has made
use of the NASA/IPAC Extragalactic Database (NED) which is operated by
the Jet Propulsion Laboratory, California Institute of Technology,
under contract with the National Aeronautics and Space Administration.


\begin{deluxetable}{llll}
\footnotesize
\tablecaption{Line IDs  \label{lineidtab}}
\tablewidth{0pt}
\tablehead{\colhead{$\lambda$ (\ang)} &\colhead{Species} &
\colhead{$\lambda$ (\ang)} &\colhead{Species}
}
\startdata
1168& N I 1168?&               1550&  C IV 1550\\                             
1668& S II 1668&	       1561&  C I 1561?\\                             
1176& C III 1176&	       1601&  Fe III 1601,1607\\                      
1192& Si II 1192, S III 1198&  1625&  Fe II 1625\\                            
1216& Ly alpha     &	       1657&  C I 1657\\                              
1234& S II 1324&	       1666&  S I 1666?, O III 1665, Al II 1671\\
1243&  N I 1243? &	       1698&   Si I 1698?\\                           
1227&  C III 1247&	       1719&   N IV 1719\\                            
1249.5& Si II 1250, 1263&      1750&   N III 1750\\                           
1252&  S II 1256&	       1805&   S II 1805\\                            
1299&  Si III 1299&	       1815&   Si II 1815\\                           
1304&  O I 1304&	       1854&   Al III 1854,1862\\                     
1335&  C II 1335&	       1892&   Si III 1982?\\                         
1338&  O IV 1338&	       1930&   C I 1930?\\                            
1342&  Si III 1342&	       2287&   Co II 2287\\                           
1346&  N II 1346?&	       2297&    C III 2297\\                          
1364.3& Si III 1364&	       2344&    Fe II 2344\\                          
1371&  O V 1371?&	       2374&    Fe II 2374\\                          
1394&  S IV 1394&	       2383&    Fe II 2383\\                          
1403&  S IV 1403&	       2396&    Fe II 2396\\                          
1428&  C III 1428?&	       2406&    Fe II 2406\\                          
1493&  N I 1493?&	       2586&    Fe II 2586,2600\\                     
1527&  Si II 1527,1533?&       2798&    Mg II h+k 2796,2804\\                 
\nodata& \nodata&	       2853&    Mg I 2853\\                           
\enddata
\end{deluxetable}

\clearpage

\begin{figure}
\epsscale{0.8}
\plotone{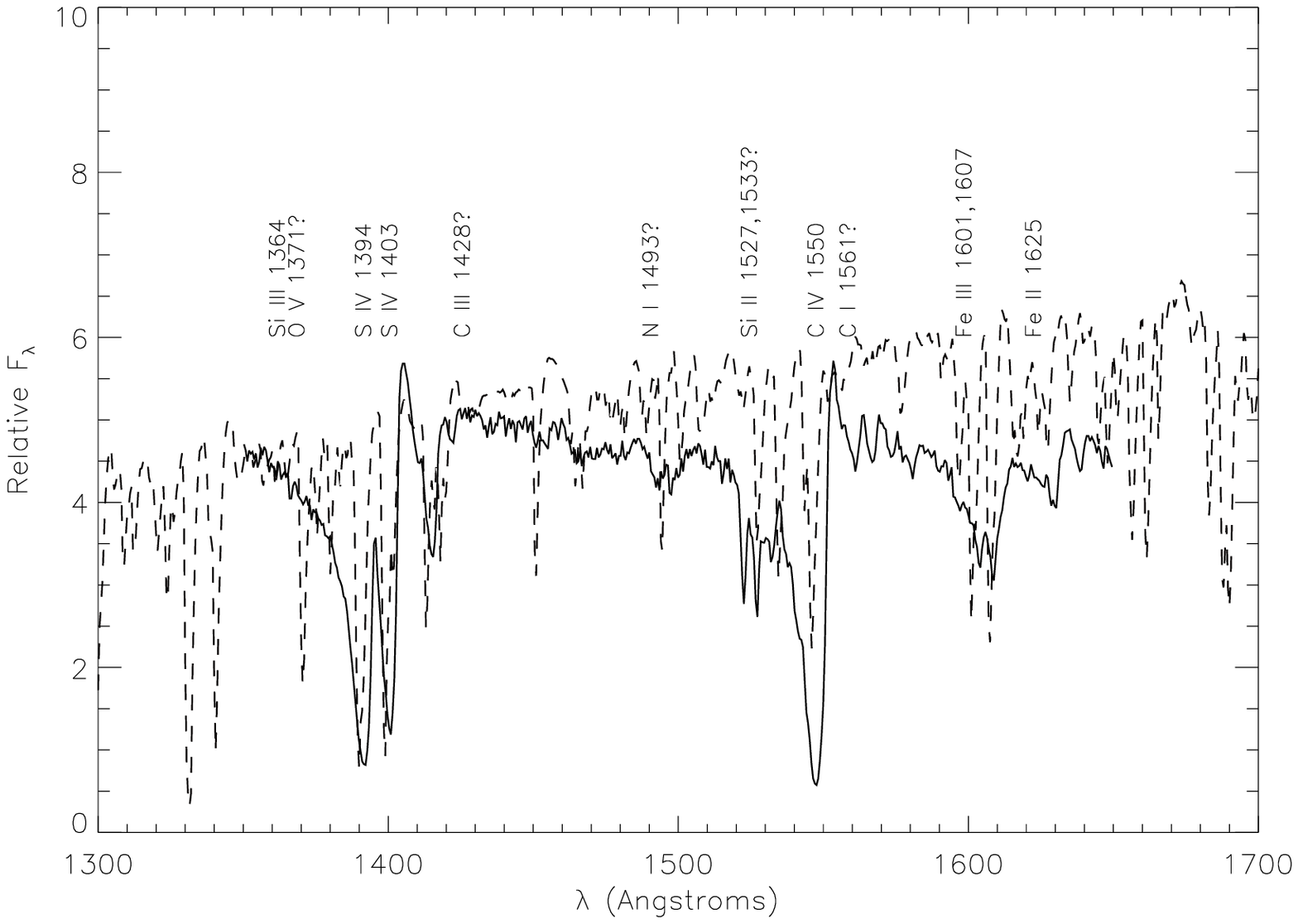}
\end{figure}

\begin{figure}
\epsscale{0.8}
\plotone{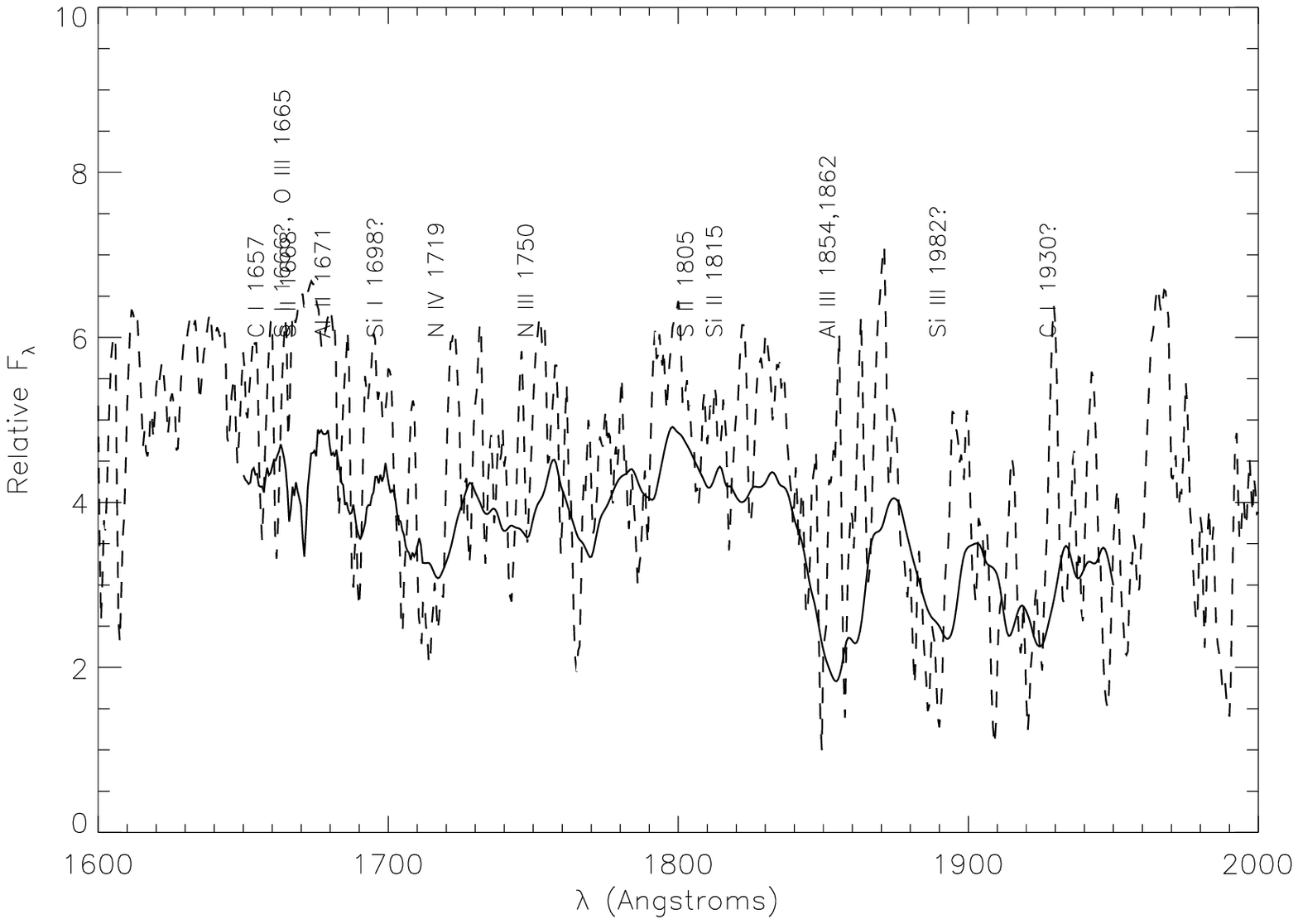}
\end{figure}

\begin{figure}
\epsscale{0.8}
\plotone{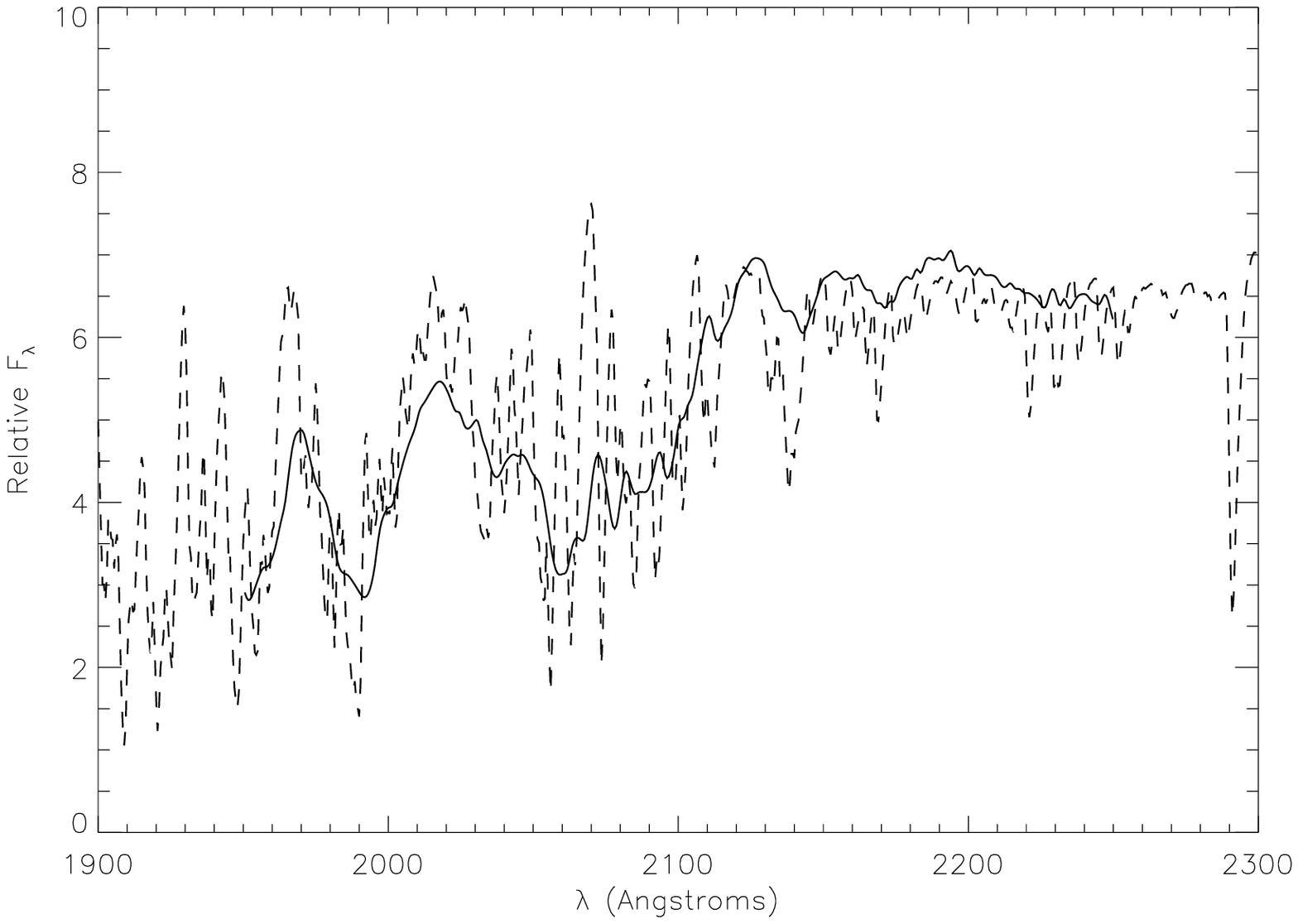}
\end{figure}

\begin{figure}
\epsscale{0.8}
\plotone{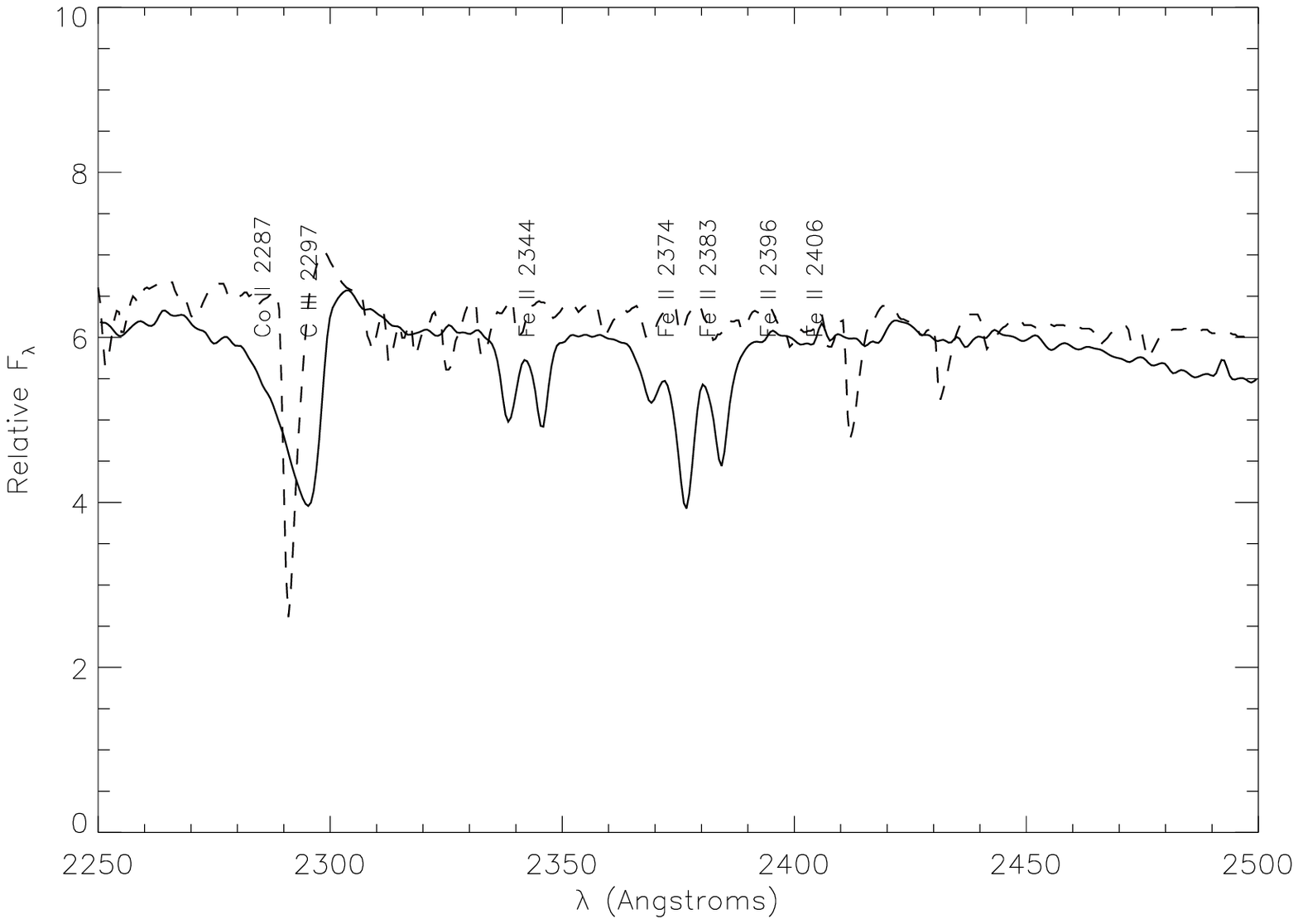}
\end{figure}

\begin{figure}
\epsscale{0.8}
\plotone{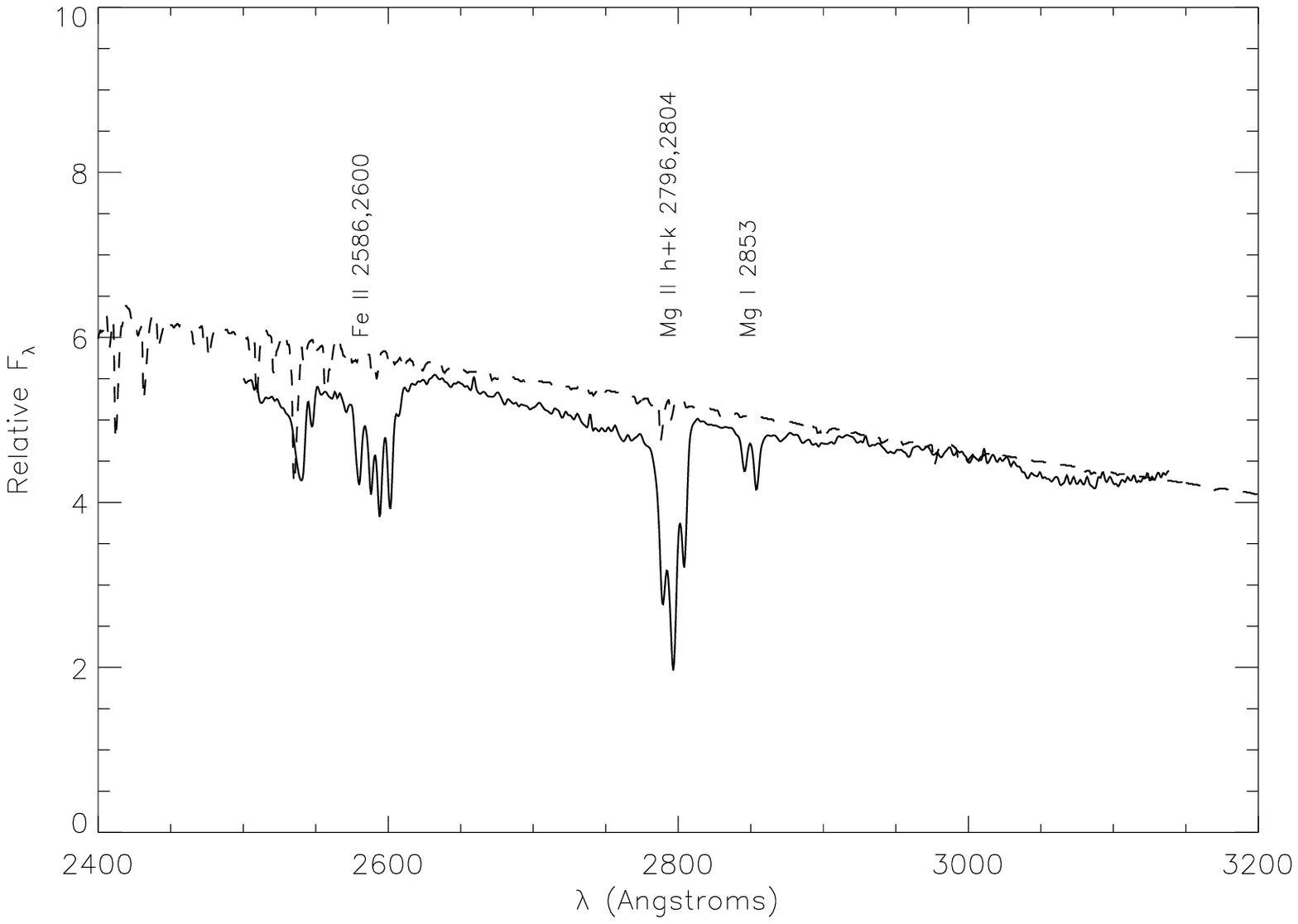}
\end{figure}

\begin{figure}
\epsscale{0.8}
\plotone{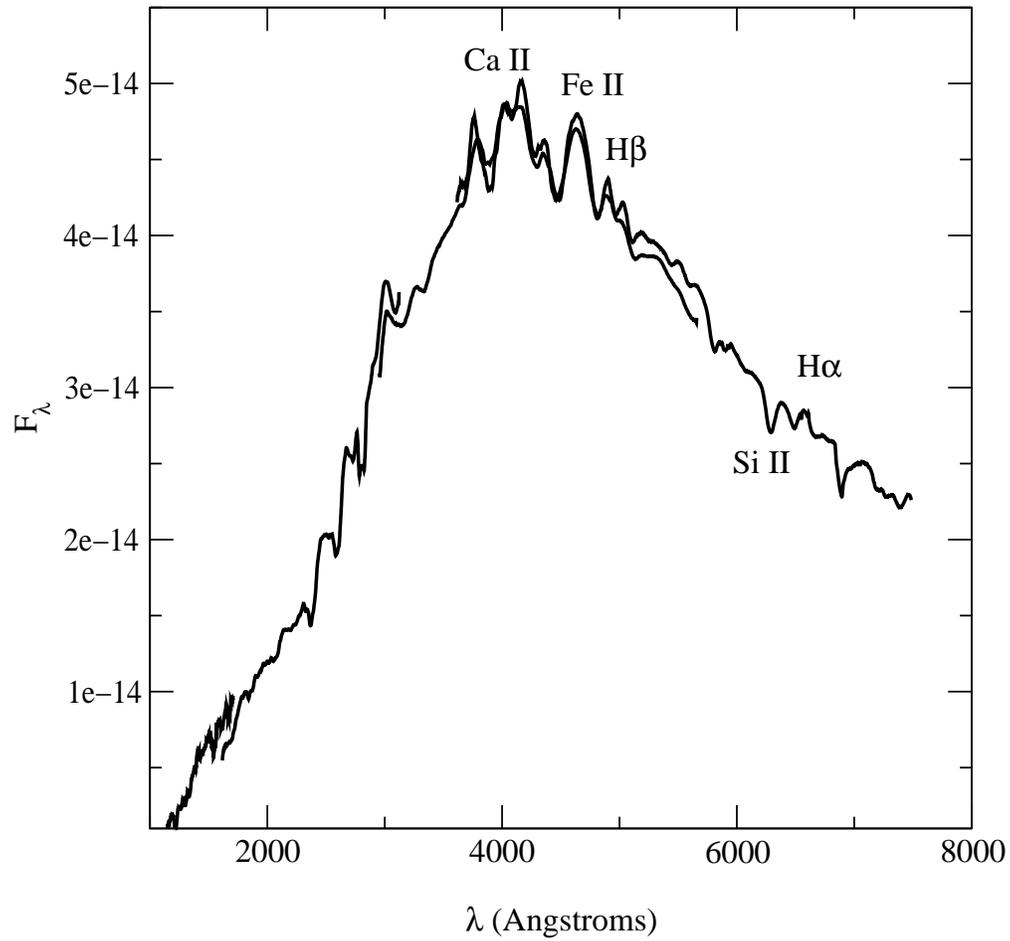}
\caption{\label{mar30data}The observed spectra from \emph{HST} and the
FLWO on Mar 30. The spectra have been smoothed using a 40 point boxcar
average, but no dereddening or deredshifting has been applied.}
\end{figure}

\begin{figure}
\epsscale{0.8}
\plotone{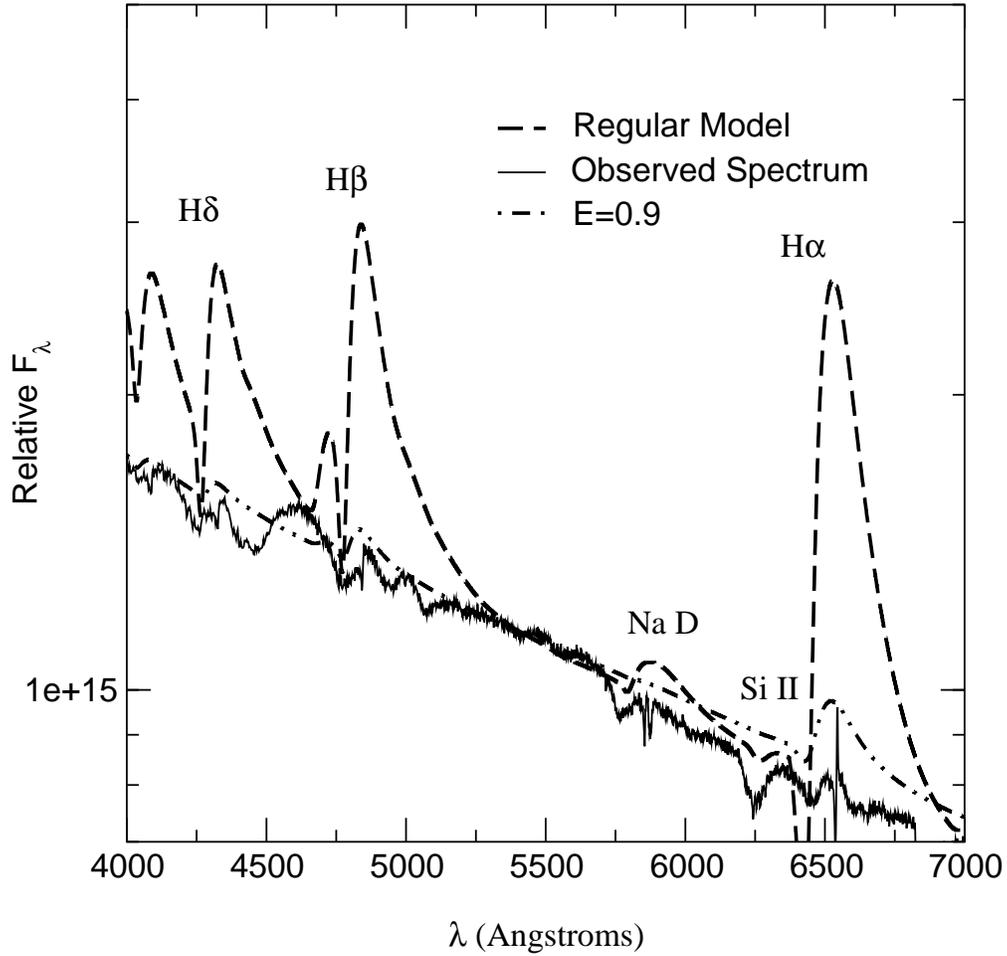}
\caption{\label{nate}Calculated synthetic spectra from the
supernova are compared with the
observed spectra in Figure~\protect\ref{mar30data}. Three spectra are
shown: the observed spectrum; a  raw ``supernova only'' synthetic
spectrum (denoted ``regular model''); and
a ``toplit'' spectrum (denoted ``$E=0.9$'').  Toplighting significantly
mutes the features, as expected \protect\cite[see][for a clear
explanation]{toplight00}. $E=0.9$  is the ratio of the CS continuum
intensity to the supernova intensity at a wavelength near H$\alpha$.}
\end{figure}

\begin{figure}
\epsscale{0.8}
\plotone{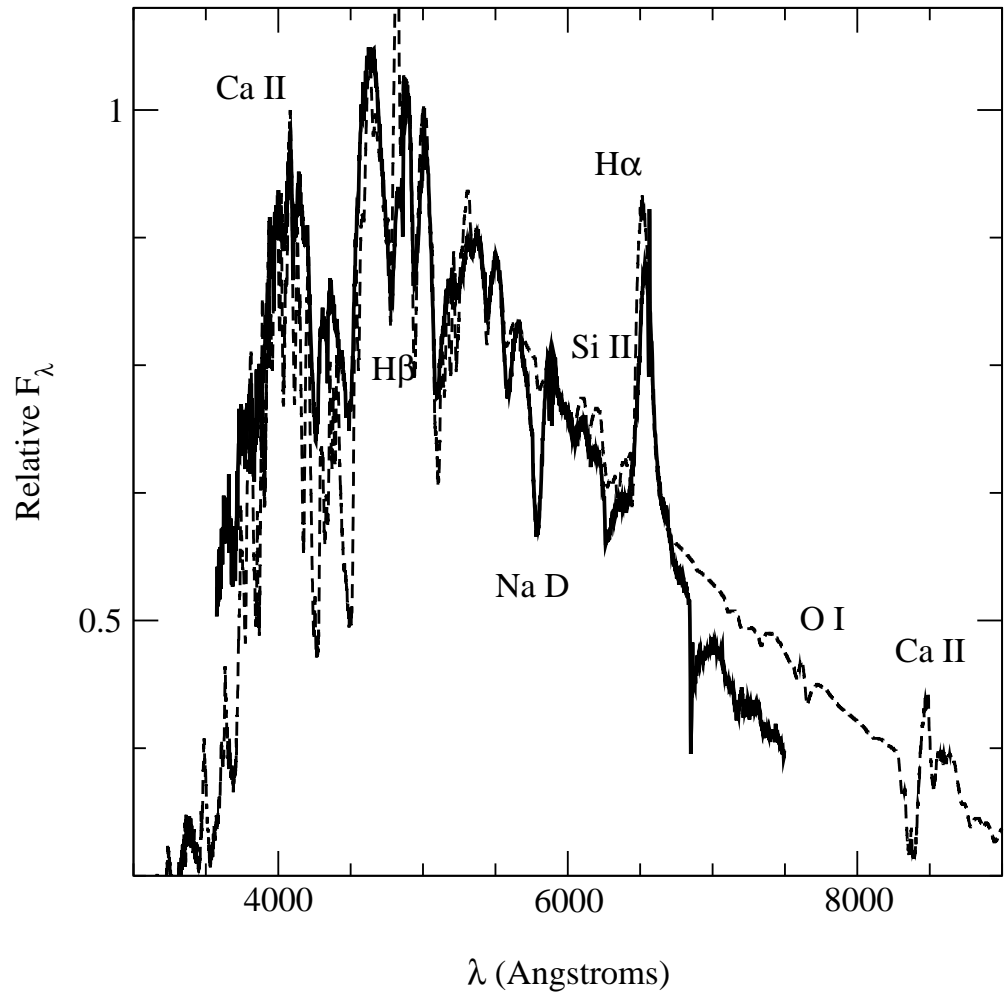}
\caption{\label{apr17opt}The calculated synthetic spectrum from the
supernova (dashed line) is compared the
optical spectrum taken at the FLWO on Apr 17, 1998.}
\end{figure}

\end{document}